\documentclass{article}
\usepackage{spconf,amsmath,graphicx}
\usepackage{hyperref}
\usepackage{subfigure}

\usepackage{amssymb}
\newcommand{\bs}[1]{\boldsymbol{#1}}

\title{STFT spectral loss for training a neural speech waveform model}
\name{Shinji Takaki$^1$, Toru Nakashika$^2$, Xin Wang$^1$, Junichi Yamagishi$^{1,3}$ \thanks{This work was partially supported by JST CREST Grant Number JPMJCR18A6, Japan and by MEXT KAKENHI Grant Numbers (16H06302, 16K16096, 17H04687, 18H04120, 18H04112, 18KT0051, 18K18069), Japan.}}
\address{$^1$National Institute of Informatics, Japan 
~~~ $^2$The University of Electro-Communications, Japan \\
$^3$The University of Edinburgh, UK \\
{\small \tt takaki@nii.ac.jp, nakashika@uec.ac.jp, wangxin@nii.ac.jp, jyamagis@nii.ac.jp}}
\begin{document}
\ninept
\maketitle
\begin{abstract}
This paper proposes a new loss using short-time Fourier transform (STFT) spectra for the aim of training a high-performance neural speech waveform model that predicts raw continuous speech waveform samples directly. Not only amplitude spectra but also phase spectra obtained from generated speech waveforms are used to calculate the proposed loss. We also mathematically show that training of the waveform model on the basis of the proposed loss can be interpreted as maximum likelihood training that assumes the amplitude and phase spectra of generated speech waveforms following Gaussian and von Mises distributions, respectively. Furthermore, this paper presents a simple network architecture as the speech waveform model, which is composed of uni-directional long short-term memories (LSTMs) and an auto-regressive structure. Experimental results showed that the proposed neural model synthesized high-quality speech waveforms.
\end{abstract}
\section{Introduction}
\label{intro}
Research on speech waveform modeling is advancing because of neural networks \cite{Takaki2017,wavenet}. The WaveNet \cite{wavenet} directly models waveform signals and demonstrates excellent performance. The WaveNet can also be used as a vocoder, which converts acoustic features, e.g., a mel-spectrogram sequence, into speech waveform signals \cite{Tamamori2017,shen2018natural}. Such a neural speech waveform model used as a vocoder can be integrated into a text-to-speech synthesis system, and it has been shown that it outperforms conventional signal-processing-based vocoders \cite{wangICASSP2018}. Neural speech waveform models will be essential components for many speech synthesis applications.

There have been several investigations of neural speech waveform models about output distribution and training criteria. A categorical distribution for discrete waveform samples and cross entropy are introduced to train the original WaveNet \cite{wavenet}. The mixture of logistic distribution and a discretized logistic mixture likelihood are also used to train the improved WaveNet \cite{oord2017parallel}. The parallel-WaveNet \cite{oord2017parallel} and the ClariNet \cite{ping2018clarinet} use a distilling approach to transfer the knowledge of auto-regressive (AR) WaveNet to a simpler non-AR student model. The distilling approach typically introduces an amplitude spectral loss as the auxiliary loss in the distillation in order to avoid generating devoiced/whisper voices. We believe that losses using short-time Fourier transform (STFT) spectra can be used beyond the distillation, and they can be used for training speech waveform models themselves. Because it is known that (complex-valued) STFT spectra represent speech characteristics well, the spectra would be useful for efficient training of neural waveform models. 

In this paper, we propose a new loss using the STFT spectra for the aim of training a high-performance neural speech waveform model that predicts raw continuous speech waveform samples directly (rather than the auxiliary loss used for the distillation process). Because not only amplitude spectra but also phase spectra represent speech characteristics \cite{interspeech2014_phase}, the proposed loss considers both amplitude and phase spectral losses obtained from generated speech waveforms. Also, we give interpretation based on probability distributions about training using the proposed loss. This interpretation provides us better understanding of the proposed method and its relationship to various spectral losses such as Kullback–-Leibler divergence or Itakura--Saito divergence \cite{lee2001algorithms,ref:Smaragdis07}. This paper also presents a simple network architecture as a speech waveform model. The proposed simple network is composed of uni-directional long short-term memories (LSTMs) and an auto-regressive structure unlike the WaveNet, which uses a relatively complicated convolutional neural network (CNN) with stacked dilated convolution.

The rest of this paper is organized as follows. Section 2 of this paper presents a neural speech waveform and the proposed loss to train it. Section 3 describes the proposed network architecture. Experimental results are presented in Section 4. We conclude in Section 5 with a summary and mention of our future work.

\section{The proposed loss for a waveform model}
A neural speech waveform model targeted in this paper is as follows.
\begin{align}
\label{eq:nn}
\bs{y} = f^{(\bs{\lambda})}(\bs{x}),
\end{align}
where, $\bs{y}\in\mathbb{R}^M$, $\bs{x}=[\bs{x}_1^{\top}, ... \bs{x}_I^{\top}]^{\top}$, $\bs{x}_i\in\mathbb{R}^D$, and $\bs{\lambda}$ represent a neural network's outputs (i.e.,\ speech waveform samples), an input sequence (e.g., a log-mel spectrogram), an input feature, and parameters of a neural network, respectively. A sample index, a frame index, and the dimension of an input feature are represented by $m$, $i$, and $D$, respectively. Back-propagation is generally used to get optimal parameters $\bs{\lambda}$ for a loss function. Because training the model is a regression task, a simple loss function is a square error between natural speech waveform samples and a neural network's outputs $\bs{y}$ as follows.
\begin{align}
\label{eq:err_wav}
E^{(wav)} = \sum_m (\hat{y}_m - y_m)^2,
\end{align}
where, $\hat{\cdot}$ denotes natural data. However, this criterion does not consider any information related to frequency characteristics of speech. In this paper, we propose loss functions using STFT spectra to effectively train a neural speech waveform instead of using the above square error as a loss function. 

\subsection{STFT spectra}
\begin{figure}[t]
  \centering
  \includegraphics[width=0.4\columnwidth]{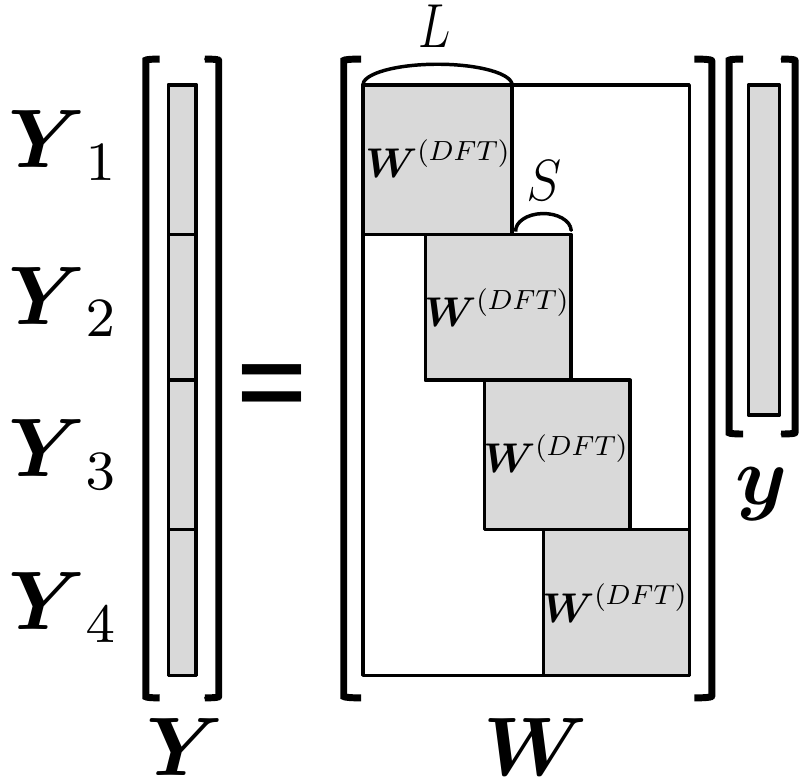}
  \vspace{-2mm}
  \caption{STFT complex spectra calculated by using a matrix $\bs{W}$. Here, $\bs{W}\in\mathbb{C}^{LT\times M}$ represents an STFT operation. $L$, $S$, and $W^{(DFT)}$ denote frame length, frame shift, and a discrete Fourier transform (DFT) matrix. White parts in the matrix $\bs{W}$ represent $0$.}
  \label{fig:STFT}
  \vspace{-5mm}
\end{figure}
In this section, we describe representations of amplitude and phase spectra used in the proposed loss. As shown in Fig.~\ref{fig:STFT}, an STFT complex spectral sequence, $\bs{Y} = [\bs{Y}_1^{\top}, ..., \bs{Y}_T^{\top}]^{\top}$, is represented by using a matrix $\bs{W}$ as follows.
\begin{align}
\label{eq:sp}
\bs{Y} = \bs{W}\bs{y},
\end{align}
where, $t$ and $\bs{W}$ represent a frame index and a matrix which performs STFT operation, respectively.
Also, a complex-valued, amplitude and phase spectra of frequency bin $n$ at frame $t$ are represented as follows.
\begin{align}
\label{eq:complex}
Y_{t,n} &= \bs{W}_{t,n} \bs{y}, \\
\label{eq:amplitude}
A_{t,n} &= |Y_{t,n}| \nonumber \\
        &= (\bs{y}^{\top}\bs{W}_{t,n}^{H}\bs{W}_{t,n} \bs{y})^{\frac{1}{2}}, \\
\label{eq:phase}
\exp(i\theta_{t,n}) &= \exp(i\angle Y_{t,n}) \nonumber \\
&= \frac{Y_{t,n}}{A_{t,n}} \nonumber \\
&= \frac{\bs{W}_{t,n} \bs{y}}{(\bs{y}^{\top}\bs{W}_{t,n}^{H}\bs{W}_{t,n} \bs{y})^{\frac{1}{2}}},
\end{align}
where, $A_{t,n}$, $\theta_{t,n}$, $\bs{W}_{t,n}$, $\cdot^{H}$, and $i$ represent an amplitude spectrum, a phase spectrum, a row vector of $\bs{W}$, Hermitian transpose, and imaginary unit, respectively. In this paper, Euler's formula is applied to a phase spectrum instead of directly using a phase spectrum. We use amplitude and phase spectra as shown in Eq.~(\ref{eq:amplitude}) and Eq.~(\ref{eq:phase}) for model training.

\vspace{-2mm}
%\subsection{Spectral losses}
\subsection{Amplitude spectral loss}
A loss function for an amplitude spectrum of frequency bin $n$ at frame $t$ is defined as a square error.
\begin{align}
\label{eq:err_amp}
E^{(amp)}_{t,n} &= \frac{1}{2}(\hat{A}_{t,n} - A_{t,n})^2 \\
&= \frac{1}{2}(\hat{A}_{t,n} - (\bs{y}^{\top}\bs{W}_{t,n}^{H}\bs{W}_{t,n} \bs{y})^{\frac{1}{2}})^2.
\end{align}
We then obtain a partial derivative of Eq.~(\ref{eq:err_amp}) w.r.t. $\bs{y}$ as
\begin{align}
\label{eq:pd_amp}
\frac{\partial{E^{(amp)}_{t,n}}}{\partial{\bs{y}}} &= \left(A_{t,n} - \hat{A}_{t,n}\right) \mathcal{R}\left(\exp(i\theta_{t,n})\bs{W}^{H}_{t,n}\right),
\end{align}
where, $\mathcal{R}(z)$ is a real part of a complex value $z$. Here the Wirtinger derivative is used to calculate the partial derivative in the complex domain. $\sum_n \partial{E^{(amp)}_{t,n}}/\partial{\bs{y}}$ can be efficiently calculated by using an inverse FFT operation.

\vspace{-2mm}
\subsection{Phase spectral loss}
A phase spectrum is a periodic variable with a period of 2$\pi$. A loss function for a phase spectrum of frequency bin $n$ at frame $t$ is defined as follows to consider this periodic property.
\begin{align}
\label{eq:err_ph}
E^{(ph)}_{t,n} &= \frac{1}{2} \left|1 - \exp(i(\hat{\theta}_{t,n}-\theta_{t,n})) \right|^2 \\
&= 1 - \frac{1}{2} \left(\frac{\hat{Y}_{t,n}}{\hat{A}_{t,n}}\frac{(\bs{y}^{\top}\bs{W}_{t,n}^{H}\bs{W}_{t,n}\bs{y})^{\frac{1}{2}}}{\bs{W}_{t,n}\bs{y}}\right.\nonumber\\
&\hspace{10mm}\left . + \frac{\overline{\hat{Y}}_{t,n}}{\hat{A}_{t,n}}\frac{(\bs{y}^{\top}\bs{W}_{t,n}^{H}\bs{W}_{t,n} \bs{y})^{\frac{1}{2}}}{\overline{\bs{W}}_{t,n}\bs{y}}\right),
\end{align}
where, $\overline{\cdot}$ is the complex conjugate. We obtain a partial derivative of Eq.~(\ref{eq:err_ph}) w.r.t. $\bs{y}$ as
\begin{align}
\label{eq:pd_ph}
\frac{\partial{E^{(ph)}_{t,n}}}{\partial{\bs{y}}} &=
\sin(\hat{\theta}_{t,n} - \theta_{t,n}) \mathcal{I}\left(\frac{1}{\overline{Y}_{t,n}}\bs{W}_{t,n}^{H}\right),
\end{align}
where, $\mathcal{I}(z)$ denotes an imaginary part of a complex value $z$. $\sum_n \partial{E^{(ph)}_{t,n}}/\partial{\bs{y}}$ can be also efficiently calculated by using an inverse FFT operation.

\vspace{-2mm}
\subsection{Loss function for model training}
A partial derivative of the amplitude loss function (Eq.~(\ref{eq:pd_amp})) includes a phase spectrum of outputs (i.e., $\theta_{t,n}$), and that of the phase loss function (Eq.~(\ref{eq:pd_ph})) includes an amplitude spectrum of outputs (i.e., $A_{t,n}$. $\overline{Y}$ can be rewritten as $A_{t,n}\exp({i(-\theta_{t,n})})$). Thus, the amplitude and phase losses are related to each other through a neural network's outputs during model training, although each loss function focuses on only amplitude or phase spectra.

In this paper, a combination of amplitude and phase spectral loss functions is used for training a neural speech waveform model.
\begin{align}
\label{eq:err}
E^{(sp)} = \sum_{t,n} (E^{(amp)}_{t,n} + \alpha_{t,n} E^{(ph)}_{t,n}),
\end{align}
where, $\alpha_{t,n}$ denotes a weight parameter. We use three types of $\alpha_{t,n}$ for training a neural speech waveform model.

\noindent
\textbf{$\alpha_{t,n} = 0$:} Using only the amplitude spectral loss function for model training.

\noindent
\textbf{$\alpha_{t,n} = 1$:} Using a simple combination of the amplitude and phase spectral loss functions.

\noindent
\textbf{$\alpha_{t,n} = v_{t}$:} Here, $v_{t}$ represents a voiced/unvoiced flag (1:voiced, 0:unvoiced). In this case, we assume that phase spectra in unvoiced parts are random values, and hence the phase spectral loss computed in unvoiced parts is omitted from model training.

\vspace{-2mm}
\subsection{Interpretation based on probability distributions}
First, Eq.~(\ref{eq:err_amp}) and Eq.~(\ref{eq:err_ph}) can be rewritten as follows.
\begin{align}
\label{eq:err_amp_dist}
E^{(amp)}_{t,n} &= \log P_{g}(\hat{A}_{t,n} \mid \hat{A}_{t,n}, 1)  - \log P_{g}(\hat{A}_{t,n} \mid A_{t,n}, 1) \\
\label{eq:err_ph_dist}
E^{(ph)}_{t,n} & = 1 - \cos(\hat{\theta}_{t,n}-\theta_{t,n})\nonumber\\
&= \log P_{vm}(\hat{\theta}_{t,n} \mid \hat{\theta}_{t,n}, 1) - \log P_{vm}(\hat{\theta}_{t,n} \mid \theta_{t,n}, 1),
\end{align} 
where, $P_{g}(\cdot)$ and $P_{vm}(\cdot)$ are probability density functions of the Gaussian distribution and the von Mises distribution.
\begin{align}
P_g(x \mid \mu, \sigma^2) &= \frac{1}{\sqrt{2\pi\sigma^2}}\exp\left(-\frac{(x-\mu)^2}{2\sigma^2}\right)  \\
P_{vm}(x \mid \mu, \beta) &= \frac{\exp(\beta\cos(x-\mu))}{2\pi I_0(\beta)}
\end{align}
$I_0(\beta)$ is the modified Bessel function of the first kind of order $0$.

From relationships of Eqs.~(\ref{eq:err_amp}), (\ref{eq:err_ph}), (\ref{eq:err_amp_dist}), and (\ref{eq:err_ph_dist}), minimization of Eq.~(\ref{eq:err}) is equivalent to that of the log negative likelihood $L$ below.
\begin{align}
\label{eq:likelihood}
L = -\log \prod_{t,n} & P_g(\hat{A}_{t,n} \mid A_{t,n}, 1)P_{vm}^{\alpha_{t,n}}(\hat{\theta}_{t,n} \mid \theta_{t,n}, 1).
\end{align}
In other words, the amplitude and phase spectra of a neural network's outputs represent mean parameters of the Gaussian distribution and the von Mises distribution, respectively. Thus, optimizing a neural network on the basis of the proposed loss function is the maximum likelihood estimation that uses the above probabilistic functions for the speech waveform model.

In this initial investigation we used the Gaussian distribution and the von Mises distribution to define loss functions for amplitude and phase spectra. But we can easily define other meaningful spectral losses by replacing the distributions with other distributions. For instance, if we use Poisson or exponential distribution instead of Gaussian distribution, we can define the Kullback–-Leibler divergence or Itakura--Saito divergence as an amplitude spectral loss \cite{lee2001algorithms,ref:Smaragdis07}. Also, we can easily define a different loss function for phase spectra by replacing the von Mises distribution with other distributions such as the generalized cardioid distribution \cite{ref:Jones05}, the generalized version of the von Mises distribution.

\section{Network architecture}
\begin{figure}[t]
  \centering
  \includegraphics[width=0.9\columnwidth]{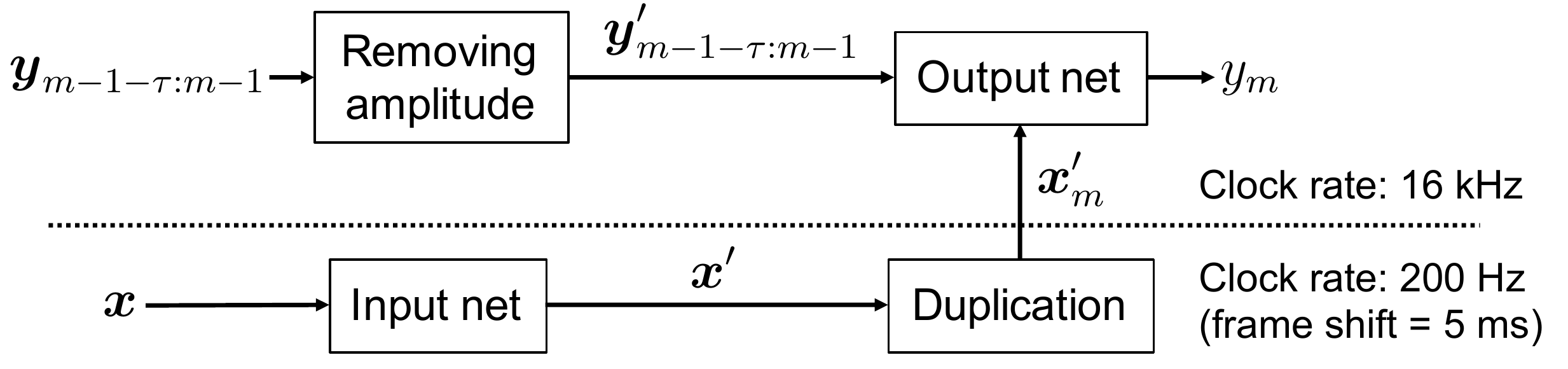}
  \caption{Overview of the proposed network.}
  \label{fig:overview}
  \vspace{-5mm}
\end{figure}
%
% %
% \begin{figure}[t]
%   \centering
%   \includegraphics[width=0.8\columnwidth]{./images/network.pdf}
%   \caption{An output network.}
%   \label{fig:output_net}
% \end{figure}
% %
Fig.~\ref{fig:overview} shows an overview of the proposed network. The processing below the dotted line in Fig.~\ref{fig:overview} is the same as that used in the WaveNet vocoder \cite{Lorenzo-Trueba2018}, in which input features are converted to hidden representations through an input network, and then they are duplicated to adjust time resolution. The processing above the dotted line is conceptually the same as that used in the WaveNet vocoder, in which feedback samples and hidden representations are fed back into an output network with an auto-regressive structure to output the next speech waveform sample. However, the following are two remarkable differences of the proposed network\footnote{We also investigated a further improved network without the auto-regressive structure in order to significantly reduce the computational cost at the synthesis phase. See our next paper \cite{ref:wang08icassp}.}.

\noindent
{\bf 1:} An output network is simply based on uni-directional LSTMs. CNNs with stacked dilated convolution are not used in our method.

\noindent
{\bf 2:} Spectral amplitude information is removed from feedback waveform samples. The output network with an auto-regressive structure feeds natural waveform samples back to the network during teacher-forced training \cite{williams1989learning}. A network composed of uni-directional LSTMs tends to rely on the feedback waveform samples while ignoring the input features if natural waveform samples are directly fed into it. To solve this problem, feedback waveform samples are converted as follows.
\begin{align}
\bs{y}'_{m-1-\tau:m-1} = \mathcal{F}^{-1} \left( \frac{\mathcal{F}(\bs{y}_{m-1-\tau:m-1})}{|\mathcal{F}(\bs{y}_{m-1-\tau:m-1})|} \right),
%% \sum_n \mathcal{R}\left(\frac{S_{n}}{|S_{n}|}\bs{W}^{(DFT),H}_{n}\right)
\end{align}
where, $\mathcal{F}$, $\mathcal{F}^{-1}$, the fraction bar, and $|\cdot|$ denote FFT operation, inverse FFT operation, element-wise division, and the element-wise absolute, respectively. Speech waveform samples, whose amplitude spectra are $1$, are obtained through the conversion. This conversion can be regarded as data dropout \cite{wang2018autoregressive}, although amplitude spectra are dropped and are replaced with $1$ instead of $0$.

\section{Experiment}
\begin{figure*}[t]
  \begin{center}
  \subfigure[NAT]{\includegraphics[height=4.8cm]{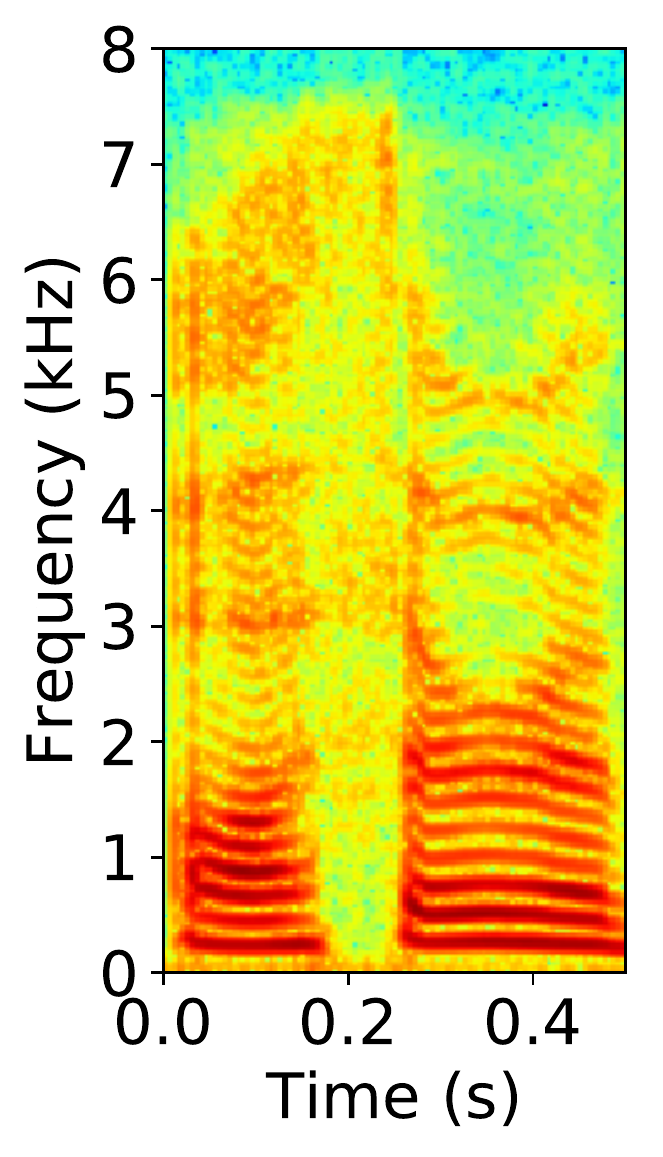}}
  \subfigure[WORLD]{\includegraphics[height=4.8cm]{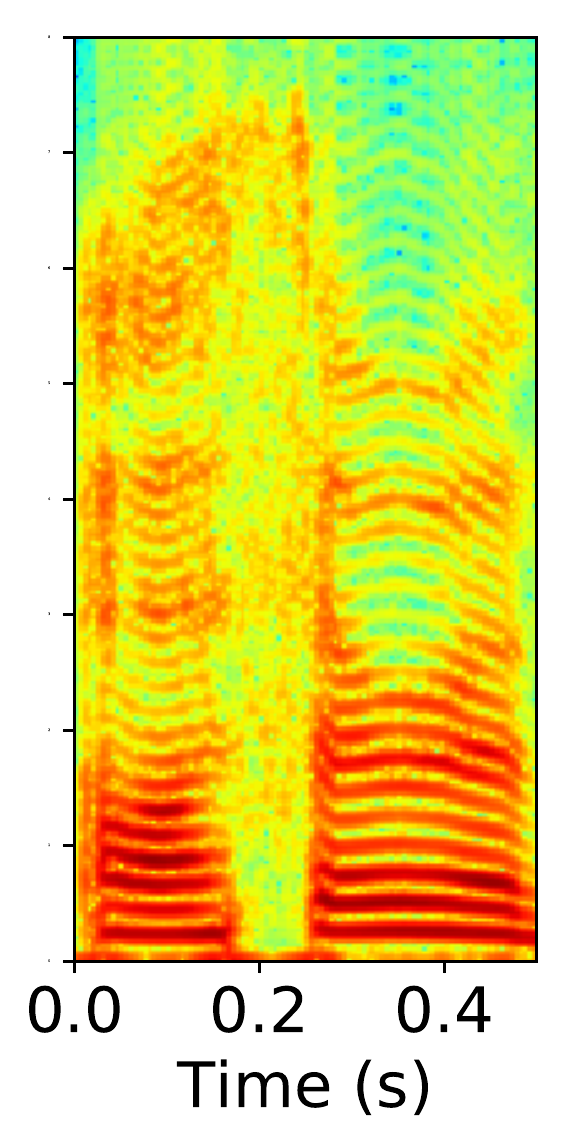}}
  \subfigure[WN]{\includegraphics[height=4.8cm]{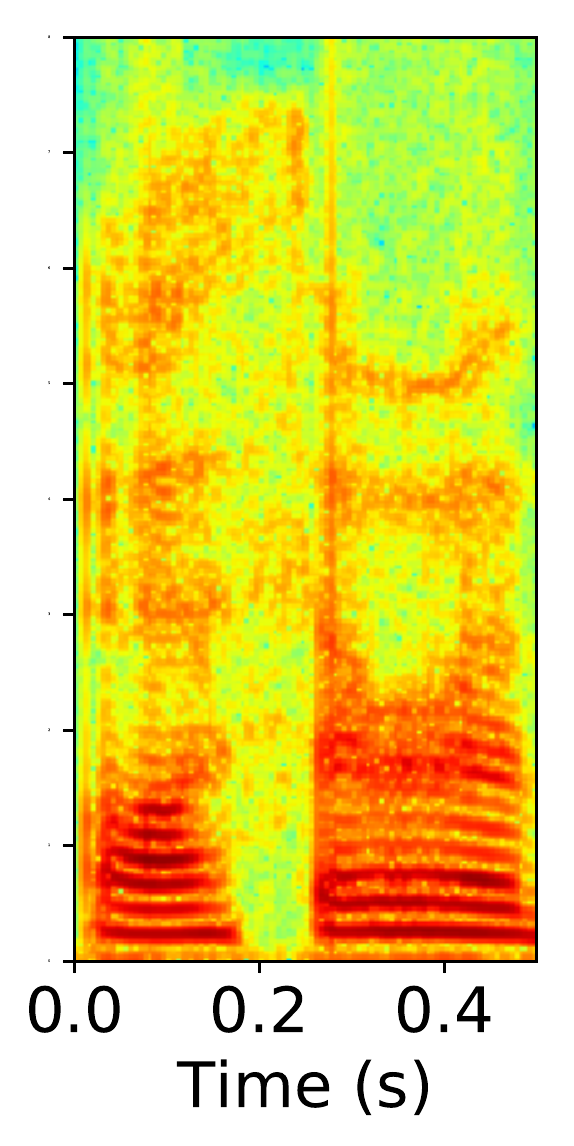}}
  \subfigure[$E^{(wav)}$]{\includegraphics[height=4.8cm]{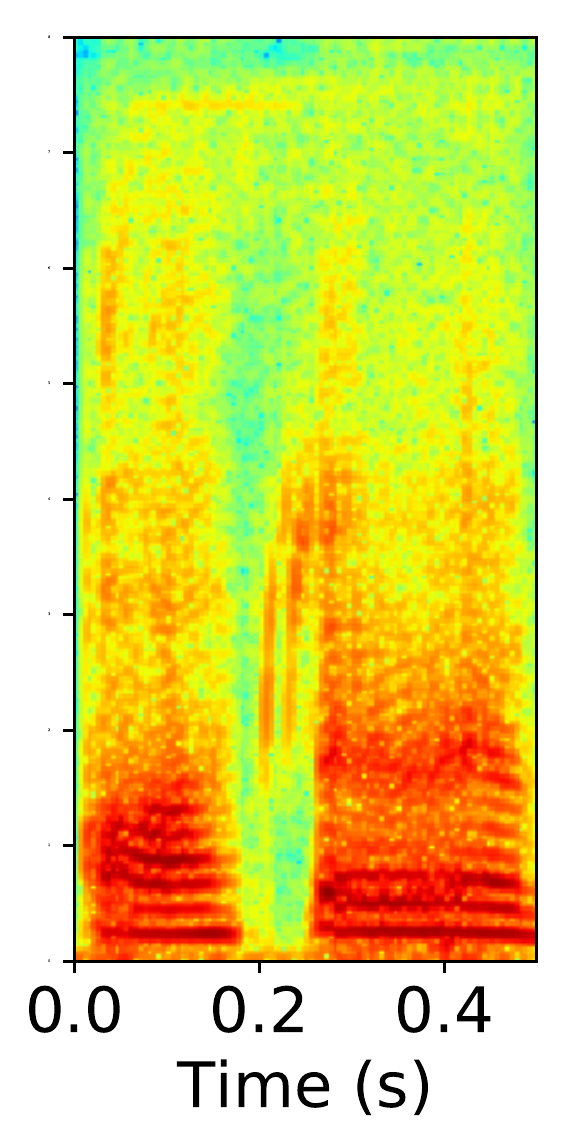}}
  \subfigure[$\alpha=0$]{\includegraphics[height=4.8cm]{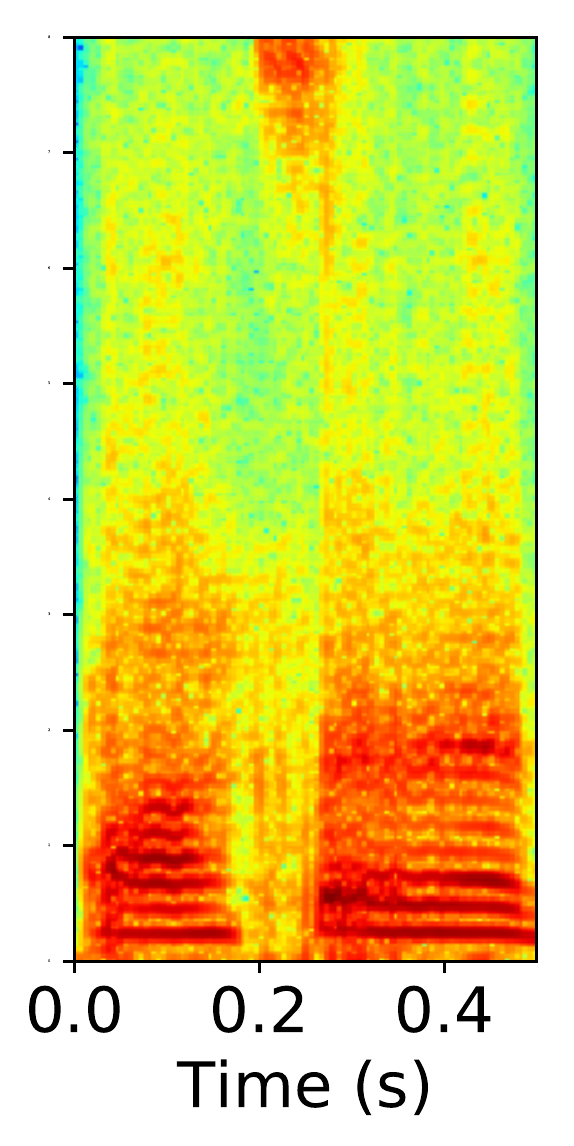}}
  \subfigure[$\alpha=1$]{\includegraphics[height=4.8cm]{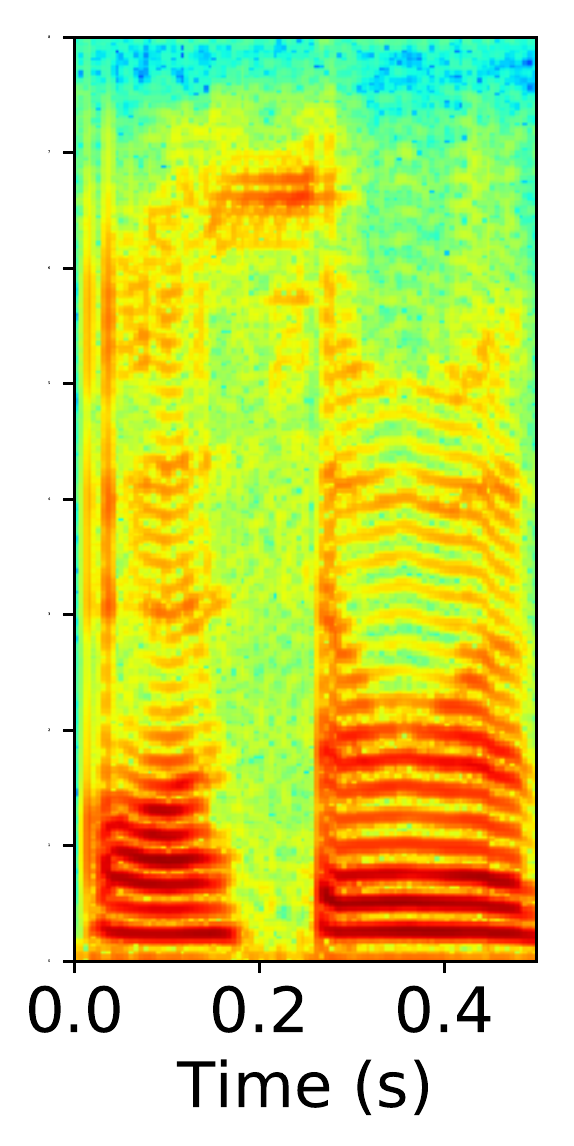}}
  \subfigure[$\alpha=v$]{\includegraphics[height=4.8cm]{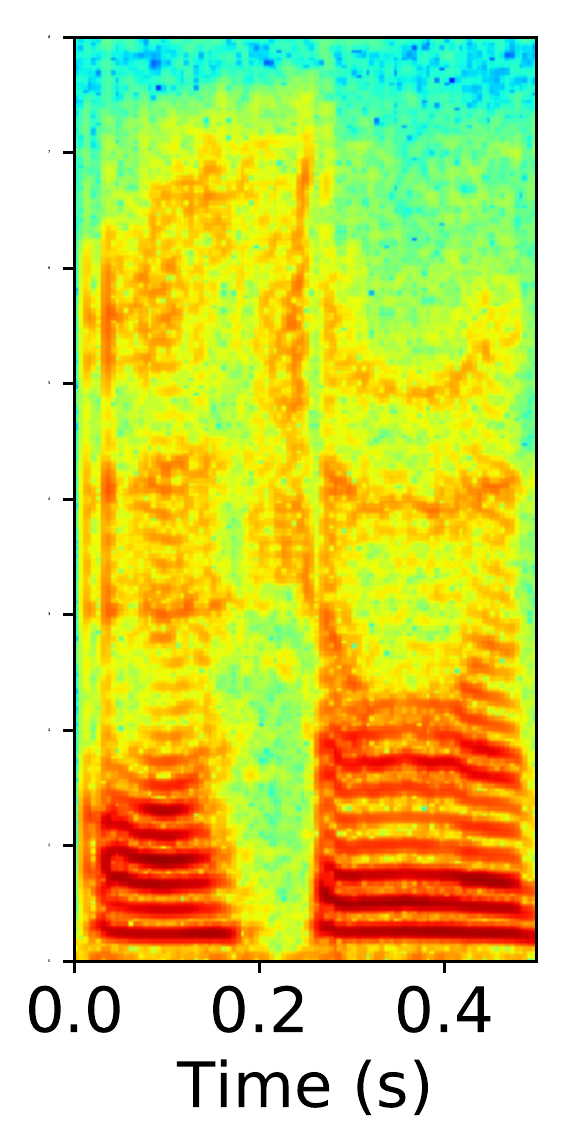}}
  \end{center}
  \vspace{-20pt}
  \caption{Spectrograms of analysis-by-synthesis speech waveform segments are shown. Here, NAT and WN are natural and the WaveNet. Spectrograms of the synthetic speech obtained by using the proposed models are shown in (d--g). Eq.~(\ref{eq:err_wav}) and Eq.~(\ref{eq:err}) with three types of $\alpha$ are used as loss functions for model training, respectively.}
  \label{fig:Spectrogram_abs}
  \vspace{-15pt}
\end{figure*}
%
%%
%\begin{figure*}[t]
%  \begin{center}
%  \subfigure[NAT]{\includegraphics[height=4.8cm]{./images/spectrogram/0NAT.pdf}}
%  \subfigure[WORLD]{\includegraphics[height=4.8cm]{./images/spectrogram/1TTS.pdf}}
%  \subfigure[WN]{\includegraphics[height=4.8cm]{./images/spectrogram/2TTS.pdf}}
%  \subfigure[$E^{(wav)}$]{\includegraphics[height=4.8cm]{./images/spectrogram/3TTS.pdf}}
%  \subfigure[$\alpha=0$]{\includegraphics[height=4.8cm]{./images/spectrogram/4TTS.pdf}}
%  \subfigure[$\alpha=1$]{\includegraphics[height=4.8cm]{./images/spectrogram/5TTS.pdf}}
%  \subfigure[$\alpha=v$]{\includegraphics[height=4.8cm]{./images/spectrogram/6TTS.pdf}}
%  \end{center}
%  \caption{Spectrograms of text-to-speech synthesis speech waveforms.}
%  \label{fig:Spectrogram_tts}
%\end{figure*}
%%

\subsection{Experimental conditions}

The proposed neural speech waveform models were evaluated as a vocoder\footnote{Synthetic speech samples and codes for model training can be found at \url{https://nii-yamagishilab.github.io/TSNetVocoder/index.html} and \url{https://github.com/nii-yamagishilab/TSNetVocoder}, respectively.}.
We used a female speaker (slt) from the CMU-ARCTIC database \cite{ref:CMUARCTIC}. $1,032$ and $50$ utterances were used as training and test sets, respectively. Their speech waveforms have a sampling frequency of 16 kHz and a 16-bit PCM format. 

As an input feature, an 80-dim log-mel spectrogram was used. Frame shift, frame length, and FFT size were $80$, $400$, and $512$, respectively. Log-mel spectrograms and waveform samples were normalized to have $0$ mean and $1$ variance for training the proposed speech waveform model.

Input features, i.e., the log-mel spectrogram sequence, were first converted into hidden representations through an input network composed of an $80$-unit bi-directional LSTM and a CNN with $80$ filters whose size is $5$ (time direction) $\times$ $80$ (frequency direction). Then, hidden representations were duplicated to adjust time resolution. The previous $400$ samples and hidden representations obtained from the input network were fed into an output network. The output network was composed of three $256$-unit Uni-directional LSTMs.

Frame shift, frame length, and FFT size used to obtain the STFT spectra of a neural network's outputs were $1$, $400$, and $512$, respectively. Mini-batches were created, each from $120$ randomly selected speech segments. Each mini-batch contained a total of 15 s speech waveform samples (each speech segment equaled 0.125 s).
We used the Adam optimizer \cite{DBLP:journals/corr/KingmaB14}, and the number of updating iteration was $100$k.

We trained four speech waveform models with the proposed network architecture. The difference among these four models was the loss function used for model training. Eq.~(\ref{eq:err_wav}) and Eq.~(\ref{eq:err}) with three types of $\alpha_{t,n}$ were used as loss functions for training the four models. For comparison, the WaveNet was trained by using $80$-dim log-mel spectrograms and 1,024 mu-law discrete waveform samples as input and output training data. The network architecture of the WaveNet was the same as that used in \cite{Lorenzo-Trueba2018}. The WORLD \cite{morise2016world} was also used as a baseline signal-processing-based vocoder. Spectral envelopes and aperiodicity measurements obtained by utilizing the WORLD were converted to 59-dim mel-cepstrum and 21-dim band aperiodicity. The total dimensions of a WORLD acoustic feature was $82$ ($60$ (mel-cepstrum) $+$ $1$ (voiced/unvoiced flag) $+$ $1$ (lf0) $+$ $21$ (band aperiodicity)). In total, we used six vocoders (the four proposed models, the WaveNet, the WORLD) in the experiment.

We evaluated analysis-by-synthesis (AbS) systems and text-to-speech (TTS) synthesis systems based on the six vocoders. For TTS synthesis systems, acoustic models that convert linguistic features to acoustic features (i.e., $80$-dim log-mel spectrogram or $82$-dim WORLD acoustic features) were separately trained. Deep auto-regressive (DAR) models \cite{wang2018autoregressive} were used as acoustic models. From the TTS experimental results, we can see the robustness of the proposed waveform model against the degraded input features generated by the acoustic models.

\vspace{-5pt}
\subsection{Experimental results}
\subsubsection{Spectrogram}
Fig.~\ref{fig:Spectrogram_abs} shows spectrograms of analysis-by-synthesis speech waveform segments. There is an unvoiced part around 0.2 seconds.
We can see from Fig.~\ref{fig:Spectrogram_abs} (d) and (e) that the proposed models trained by using $E^{(wav)}$ and $E^{(sp)}$ without the phase spectral loss function (i.e., $\alpha=0$) generated noisy spectrograms. Also, it can be seen from Fig.~\ref{fig:Spectrogram_abs} (f) that large value amplitude spectra are observed around $7$ kHz in the unvoiced part, whereas they are not observed in the natural spectrogram. Such artifacts are not observed in the spectrogram shown as Fig.~\ref{fig:Spectrogram_abs} (g). These results indicate that training a proposed model by using the amplitude and phase spectral loss functions is adequate and adjusting the weight parameter $\alpha$ by using the voiced/unvoiced flag further improves the performance.

Both spectrograms of the WaveNet and the proposed model trained by adjusting $\alpha_{t,n}$ by using the voiced/unvoiced flags (i.e., Fig.~\ref{fig:Spectrogram_abs} (c) and (g)) are similar to the natural one, although the detailed spectral structures are changed.

\subsubsection{Subjective evaluation result}
\begin{figure}[t]
  \centering
  \includegraphics[width=0.9\columnwidth]{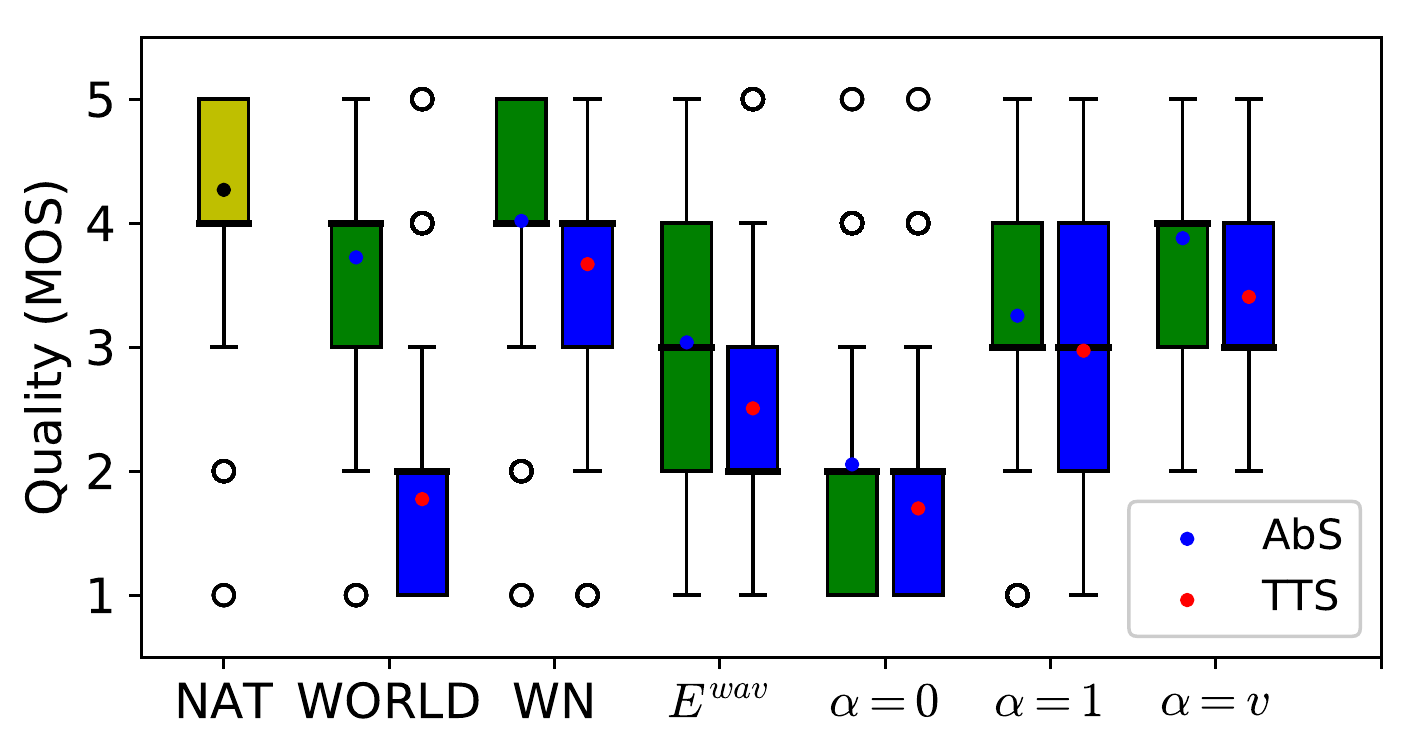}
    \vspace{-5mm}  
  \caption{Box plots on naturalness evaluation results. Blue and red dots represent the mean results of analysis-by-synthesis and text-to-speech synthesis systems, respectively.}
  \label{fig:naturalness}
  \vspace{-4mm}
\end{figure}
The subjective evaluation was conducted using 158 crowd-sourced listeners. Natural samples and samples synthesized from the 12 systems were evaluated. The number of synthetic samples was $650$ ($13$ systems $\times$ $50$ test sentences). Participants evaluated speech naturalness on a five-point mean opinion score (MOS) scale. Each synthetic sample was evaluated $40$ times, giving a total of $26,000$ data points. Thanks to the large number of data points, differences between all combinations of evaluated systems are statistically significant (i.e., $p< 0.05$).

First, among the proposed models, those trained together with the phase spectral loss outperformed the others. Using not only amplitude spectra but also phase spectra to calculate loss is useful for training a neural speech waveform model. The proposed model trained by adjusting $\alpha_{t,n}$ by using the voiced/unvoiced flags obtained the best score among the proposed models.

Second, it can be seen from Fig.~\ref{fig:naturalness} that the best performance of the proposed model (i.e., $\alpha = v$) is better than that of the WORLD. In the WORLD, the TTS result is drastically decreased from the AbS result. On the other hand, we can see that the proposed model with the best configuration (i.e., $\alpha = v$) is robust against acoustic features predicted from the acoustic model.

Finally, compared with the WaveNet, the best performance of the proposed models (i.e., $\alpha = v$) is slightly worse. But note that the number of parameters of the proposed model (1.8M) is less than that of the WaveNet (2.3M). Although we need to improve loss functions and network architecture to achieve comparable performance with the WaveNet, it is obvious that the proposed model achieved high-performance speech waveform modeling.

\vspace{-2mm}
\section{Conclusion}
\vspace{-2mm}
We proposed a new STFT spectral loss to train a high-performance speech waveform model directly. We also presented a simple network architecture for the model, which is composed of uni-directional LSTMs and an auto-regressive structure. Experimental results showed that the proposed model can synthesize high-quality speech waveforms. 

This is a part of our sequential work. In our next paper \cite{ref:wang08icassp}, we will show that the above model can be improved further to achieve performance comparable with the WaveNet. Our other future work includes training based on other time-frequency analysis such as modified discrete cosine transform.

\vfill\pagebreak

% References should be produced using the bibtex program from suitable
% BiBTeX files (here: strings, refs, manuals). The IEEEbib.bst bibliography
% style file from IEEE produces unsorted bibliography list.
% -------------------------------------------------------------------------

\bibliographystyle{IEEEbib}
\bibliography{refs}

\begin{thebibliography}{10}

\bibitem{Takaki2017}
Shinji Takaki, Hirokazu Kameoka, and Junichi Yamagishi,
\newblock ``Direct modeling of frequency spectra and waveform generation based
  on phase recovery for {DNN}-based speech synthesis,''
\newblock in {\em Proc. Interspeech}, 2017, pp. 1128--1132.

\bibitem{wavenet}
A{\"{a}}ron van~den Oord, Sander Dieleman, Heiga Zen, Karen Simonyan, Oriol
  Vinyals, Alex Graves, Nal Kalchbrenner, Andrew~W. Senior, and Koray
  Kavukcuoglu,
\newblock ``Wavenet: {A} generative model for raw audio,''
\newblock {\em CoRR}, vol. abs/1609.03499, 2016.

\bibitem{Tamamori2017}
Akira Tamamori, Tomoki Hayashi, Kazuhiro Kobayashi, Kazuya Takeda, and Tomoki
  Toda,
\newblock ``Speaker-dependent {WaveNet} vocoder,''
\newblock in {\em Proc. Interspeech}, 2017, pp. 1118--1122.

\bibitem{shen2018natural}
Jonathan Shen, Ruoming Pang, Ron~J Weiss, Mike Schuster, Navdeep Jaitly,
  Zongheng Yang, Zhifeng Chen, Yu~Zhang, Yuxuan Wang, Rj~Skerrv-Ryan, et~al.,
\newblock ``Natural {TTS} synthesis by conditioning {WaveNet} on {Mel}
  spectrogram predictions,''
\newblock in {\em Proc. ICASSP}, 2018, pp. 4779--4783.

\bibitem{wangICASSP2018}
Xin Wang, Jaime Lorenzo-Trueba, Shinji Takaki, Lauri Juvela, and Junichi
  Yamagishi,
\newblock ``A comparison of recent waveform generation and acoustic modeling
  methods for neural-network-based speech synthesis,''
\newblock in {\em Proc. ICASSP}, 2018, pp. 4804--4808.

\bibitem{oord2017parallel}
Aaron van~den Oord, Yazhe Li, Igor Babuschkin, Karen Simonyan, Oriol Vinyals,
  Koray Kavukcuoglu, George van~den Driessche, Edward Lockhart, Luis~C Cobo,
  Florian Stimberg, et~al.,
\newblock ``Parallel {WaveNet}: Fast high-fidelity speech synthesis,''
\newblock {\em arXiv preprint arXiv:1711.10433}, 2017.

\bibitem{ping2018clarinet}
Wei Ping, Kainan Peng, and Jitong Chen,
\newblock ``Clarinet: Parallel wave generation in end-to-end text-to-speech,''
\newblock {\em arXiv preprint arXiv:1807.07281}, 2018.

\bibitem{interspeech2014_phase}
Pejman Mowlaee, Rahim Saeidi, and Yannis Stylianou,
\newblock ``{INTERSPEECH} 2014 special session: Phase importance in speech
  processing applications,''
\newblock {\em Interspeech}, pp. 1623--1627, 2014.

\bibitem{lee2001algorithms}
Daniel~D Lee and H~Sebastian Seung,
\newblock ``Algorithms for non-negative matrix factorization,''
\newblock in {\em Proc. NIPS}, 2001, pp. 556--562.

\bibitem{ref:Smaragdis07}
P.~Smaragdis, B.~Raj, and M.~Shashanka,
\newblock ``Supervised and semi-supervised separation of sounds from
  single-channel mixtures,''
\newblock {\em Proceedings of 7th Int. Conf. Ind. Compon. Anal. Signal
  Separat.}, pp. 414–--421, 2007.

\bibitem{ref:Jones05}
M.~C. Jones and Arthur Pewsey,
\newblock ``A family of symmetric distributions on the circle,''
\newblock {\em Journal of the American Statistical Association}, vol. 100, no.
  472, pp. 1422–--1428, 2005.

\bibitem{Lorenzo-Trueba2018}
Jaime Lorenzo-Trueba, Fuming Fang, Xin Wang, Isao Echizen, Junichi Yamagishi,
  and Tomi Kinnunen,
\newblock ``Can we steal your vocal identity from the internet?: Initial
  investigation of cloning obama’s voice using gan, wavenet and low-quality
  found data,''
\newblock in {\em Proc. Odyssey 2018 The Speaker and Language Recognition
  Workshop}, 2018, pp. 240--247.

\bibitem{ref:wang08icassp}
Xin Wang, Shinji Takaki, and Junichi Yamagishi,
\newblock ``Neural source-filter-based waveform model for statistical
  parametric speech synthesis,''
\newblock {\em Submitted to ICASSP 2019}, 2019.

\bibitem{williams1989learning}
Ronald~J Williams and David Zipser,
\newblock ``A learning algorithm for continually running fully recurrent neural
  networks,''
\newblock {\em Neural computation}, vol. 1, no. 2, pp. 270--280, 1989.

\bibitem{wang2018autoregressive}
Xin Wang, Shinji Takaki, and Junichi Yamagishi,
\newblock ``Autoregressive neural f0 model for statistical parametric speech
  synthesis,''
\newblock {\em IEEE/ACM Transactions on Audio, Speech, and Language
  Processing}, vol. 26, no. 8, pp. 1406--1419, 2018.

\bibitem{ref:CMUARCTIC}
J.~Kominek and A.~W. Black,
\newblock ``The {CMU} arctic speech databases,''
\newblock {\em Fifth ISCA Workshop on Speech Synthesis}, 2004.

\bibitem{DBLP:journals/corr/KingmaB14}
Diederik~P. Kingma and Jimmy Ba,
\newblock ``Adam: {A} method for stochastic optimization,''
\newblock {\em CoRR}, vol. abs/1412.6980, 2014.

\bibitem{morise2016world}
Masanori Morise, Fumiya Yokomori, and Kenji Ozawa,
\newblock ``{WORLD}: A vocoder-based high-quality speech synthesis system for
  real-time applications,''
\newblock {\em IEICE Trans. on Information and Systems}, vol. 99, no. 7, pp.
  1877--1884, 2016.

\end{thebibliography}

\clearpage
\onecolumn
\appendix
\allowdisplaybreaks[1]
\section{Details of the partial derivatives}
The Wirtinger derivative is used to calculate the partial derivative with respect to a complex value $z$ in the complex domain as,
\begin{align}
\frac{d E}{d z} &= \frac{1}{2}\left(\frac{\partial{E}}{\partial{\mathcal{R}(z)}} - i \frac{\partial{E}}{\partial{\mathcal{I}(z)}} \right), \\
\frac{d E}{d \overline{z}} &= \frac{1}{2}\left(\frac{\partial{E}}{\partial{\mathcal{R}(z)}} + i \frac{\partial{E}}{\partial{\mathcal{I}(z)}} \right).
\end{align}
If $E$ is a real function, the complex gradient vector is given by
\begin{align}
\nabla E &= 2 \frac{d E}{d \overline{\bs{z}}} \\
&= \frac{\partial{E}}{\partial{\mathcal{R}(\bs{z})}} + i \frac{\partial{E}}{\partial{\mathcal{I}(\bs{z})}}.
\end{align}
For a non-analytic function, the chain rule is given by
\begin{align}
\frac{\partial{E}}{\partial{x}} = \frac{\partial{E}}{\partial{u}}\frac{\partial{u}}{\partial{x}} + \frac{\partial{E}}{\partial{\overline{u}}}\frac{\partial{\overline{u}}}{\partial{x}}.
\end{align}

\subsection{Derivative of amplitude spectral loss $E^{(amp)}_{t,n}$}
Given the amplitude spectral loss
\begin{align}
E^{(amp)}_{t,n} = \frac{1}{2}(\hat{A}_{t,n} - A_{t,n})^2.
\end{align}
According to the chain rule, we can compute the derivative
\begin{equation}
\frac{\partial{E^{(amp)}_{t,n}}}{\partial{\bs{y}}} = ({A}_{t,n} - \hat{A}_{t,n}) \cdot \frac{\partial{{A}_{t,n}}}{\partial{\bs{y}}}.
\label{eq:append_b_amp_error_0}
\end{equation}
For $\frac{\partial{{A}_{t,n}}}{\partial{\bs{y}}}$, since we know that $A_{t,n} = (\bs{y}^{\top}\bs{W}_{t,n}^{H}\bs{W}_{t,n} \bs{y})^{\frac{1}{2}}\in\mathbb{R}$, we can compute
\begin{align}
\frac{\partial{A_{t,n}}}{\partial{\bs{y}}} &= \frac{\partial{(\bs{y}^{\top}\bs{W}_{t,n}^{H}\bs{W}_{t,n} \bs{y})^{\frac{1}{2}}}}{\partial{\bs{y}}} \label{eq:append_b_amp_error_1}\\
&= \frac{1}{2}(\bs{y}^{\top}\bs{W}_{t,n}^{H}\bs{W}_{t,n} \bs{y})^{-\frac{1}{2}} \cdot (\bs{W}_{t,n}^{H}\bs{W}_{t,n} + \bs{W}_{t,n}^{\top}\overline{\bs{W}}_{t,n})\bs{y} 
\label{eq:append_b_amp_error_2} \\
&= \frac{1}{2}(\bs{y}^{\top}\bs{W}_{t,n}^{H}\bs{W}_{t,n} \bs{y})^{-\frac{1}{2}} \cdot (\bs{W}_{t,n}^{H}\bs{W}_{t,n}\bs{y} + \bs{W}_{t,n}^{\top}\overline{\bs{W}}_{t,n}\bs{y}) \label{eq:append_b_amp_error_3}\\
&= \frac{1}{2}(\bs{y}^{\top}\bs{W}_{t,n}^{H}\bs{W}_{t,n} \bs{y})^{-\frac{1}{2}} \cdot 2\mathcal{R}(\bs{W}_{t,n}^{H}\bs{W}_{t,n}\bs{y})  
\label{eq:append_b_amp_error_4}\\
&= \frac{1}{A_{t,n}} \cdot \mathcal{R}(Y_{t,n}\bs{W}_{t,n}^{H})
\label{eq:append_b_amp_error_5} \\
&= \mathcal{R}(\frac{Y_{t,n}}{A_{t,n}}\bs{W}_{t,n}^{H})
\label{eq:append_b_amp_error_6} \\
&= \mathcal{R}(\exp({i\theta_{t,n}})\bs{W}_{t,n}^{H})
\label{eq:append_b_amp_error_7}
\end{align}
where, $\cdot^{\top}$ and $\cdot^{H}$ denotes transpose and Hermititian transpose, respectively. 
Note that, from Eq.~(\ref{eq:append_b_amp_error_1}) to ~(\ref{eq:append_b_amp_error_2}), because $W_{t,n}$ is a complex-valued vector, we need to use  
\begin{equation}
\frac{\partial{\bs{y}^{\top}\bs{W}_{t,n}^{H}\bs{W}_{t,n} \bs{y}}}{\partial{\bs{y}}} = (\bs{W}_{t,n}^{H}\bs{W}_{t,n} + \bs{W}_{t,n}^{\top}\overline{\bs{W}}_{t,n})\bs{y}.
\end{equation}
From Eqs.~(\ref{eq:append_b_amp_error_3}) to ~(\ref{eq:append_b_amp_error_4}), we use the fact that $\overline{\bs{W}_{t,n}^{H}\bs{W}_{t,n}\bs{y}} = {\bs{W}_{t,n}^{\top}\overline{\bs{W}}_{t,n}\bs{y}}$ and therefore
\begin{equation}
{\bs{W}_{t,n}^{H}\bs{W}_{t,n}\bs{y}} + {\bs{W}_{t,n}^{\top}\overline{\bs{W}}_{t,n}\bs{y}} = 2\mathcal{R}(\bs{W}_{t,n}^{H}\bs{W}_{t,n}\bs{y}).
\end{equation}
Based on Equation~\ref{eq:append_b_amp_error_7} and \ref{eq:append_b_amp_error_0}, we finally get
\begin{equation}
\frac{\partial{E^{(amp)}_{t,n}}}{\partial{\bs{y}}} = (A_{t,n} - \hat{A}_{t,n})\mathcal{R}(\exp(i\theta_{t,n})\bs{W}_{t,n}^{H})).
\label{eq:append_b_amp_error_8}
\end{equation}

\subsection{Derivative of phase spectrum loss $E^{(ph)}_{t,n}$}
Based on the definition of $E^{(ph)}_{t,n}$, we get
\begin{align}
E^{(ph)}_{t,n} &= \frac{1}{2} \left|1 - \exp(i(\hat{\theta}_{t,n}-\theta_{t,n})) \right|^2 \label{eq:append_b_ph_1}\\
&= \frac{1}{2} \left|1 - \exp(i(\hat{\theta}_{t,n}))\frac{1}{\exp(i(\theta_{t,n}))} \right|^2 \label{eq:append_b_ph_2}\\
&= \frac{1}{2} \left|1 - \frac{\hat{Y}_{t,n}}{\hat{A}_{t,n}}\frac{{A_{t,n}}}{{Y_{t,n}}} \right|^2 \label{eq:append_b_ph_3}\\
&= \frac{1}{2} \left(1 - \frac{\hat{Y}_{t,n}}{\hat{A}_{t,n}}\frac{{A_{t,n}}}{{Y_{t,n}}} \right) \overline{\left(1 - \frac{\hat{Y}_{t,n}}{\hat{A}_{t,n}}\frac{{A_{t,n}}}{{Y_{t,n}}} \right)} \label{eq:append_b_ph_4}\\
&= \frac{1}{2} \left(1 - \frac{\hat{Y}_{t,n}}{\hat{A}_{t,n}}\frac{{A_{t,n}}}{{Y_{t,n}}} \right)\left(1 - \frac{\overline{\hat{Y}}_{t,n}}{\hat{A}_{t,n}}\frac{{A_{t,n}}}{{\overline{Y}_{t,n}}} \right) \label{eq:append_b_ph_5}\\
&= \frac{1}{2} \left(1 + \frac{\hat{Y}_{t,n}}{\hat{A}_{t,n}}\frac{{A_{t,n}}}{{Y_{t,n}}}\frac{\overline{\hat{Y}}_{t,n}}{\hat{A}_{t,n}}\frac{{A_{t,n}}}{{\overline{Y}_{t,n}}} - \left(\frac{\hat{Y}_{t,n}}{\hat{A}_{t,n}}\frac{{A_{t,n}}}{{Y_{t,n}}}+\frac{\overline{\hat{Y}}_{t,n}}{\hat{A}_{t,n}}\frac{{A_{t,n}}}{{\overline{Y}_{t,n}}}\right)\right) \label{eq:append_b_ph_6}\\
&= \frac{1}{2} \left(1 + 1 - \left(\frac{\hat{Y}_{t,n}}{\hat{A}_{t,n}}\frac{{A_{t,n}}}{{Y_{t,n}}}+\frac{\overline{\hat{Y}}_{t,n}}{\hat{A}_{t,n}}{\frac{{A_{t,n}}}{\overline{Y}_{t,n}}}\right)\right) \label{eq:append_b_ph_7}\\
&= 1 - \frac{1}{2}\left(\frac{\hat{Y}_{t,n}}{\hat{A}_{t,n}}\frac{{A_{t,n}}}{{Y_{t,n}}}+\frac{\overline{\hat{Y}}_{t,n}}{\hat{A}_{t,n}}{\frac{{A_{t,n}}}{\overline{Y}_{t,n}}}\right) \\
&= 1 - \frac{1}{2}\frac{1}{\hat{A}_{t,n}A_{t,n}}\left(\overline{\hat{Y}}Y + \hat{Y}\overline{Y}\right).
\label{eq:append_b_ph_8}
\end{align}
Note that $\hat{A}_{t,n}$, ${A}_{t,n}$ and $\hat{Y}\overline{Y}+\overline{\hat{Y}}Y\in\mathbb{R}$ while $\hat{Y}_{t,n}$ and ${Y}_{t,n}\in\mathbb{C}$. Also note that $\overline{{Y}}_{t,n}{Y}_{t,n}={A}_{t,n}^2$ and $\overline{\hat{Y}}_{t,n}\hat{Y}_{t,n}=\hat{A}_{t,n}^2$.
% Also, $E^{(ph)}_{t,n}$ can be rewritten as,
% \begin{align}
% E^{(ph)}_{t,n} &= 1 - \frac{1}{2}\left(\frac{\hat{Y}_{t,n}}{\hat{A}_{t,n}}\frac{{A_{t,n}}}{{Y_{t,n}}} + \frac{\overline{\hat{Y}}_{t,n}}{\hat{A}_{t,n}} \frac{A_{t,n}}{\overline{Y}_{t,n}} \right) \\
% &= 1 - \frac{1}{2}(\exp(i(\hat{\theta}_{t,n} - \theta{t,n})) + \exp(i(-(\hat{\theta}_{t,n} - \theta_{t,n})))) \\
% &= 1 - \cos(\hat{\theta}_{t,n} - \theta_{t,n})
% \end{align}
Thus, we can calculate $\frac{\partial{E^{(ph)}_{t,n}}}{\partial{\bs{y}}}$ as follows.
\begin{align}
\frac{\partial{E^{(ph)}_{t,n}}}{\partial{\bs{y}}} &= \frac{1}{2}\left(\frac{\partial{\hat{A}_{t,n}^{-1}A_{t,n}^{-1}}}{\partial{\bs{y}}}\left(\overline{\hat{Y}}_{t,n}Y_{t,n} + \hat{Y}_{t,n}\overline{Y}_{t,n}\right) + \frac{1}{\hat{A}_{t,n}A_{t,n}}\frac{\partial \overline{\hat{Y}}_{t,n}Y_{t,n} + \hat{Y}_{t,n}\overline{Y}_{t,n}}{\partial{\bs{y}}} \right) \label{eq:cos_pd_1} \\
&=\frac{1}{2}\left(-\frac{1}{2\hat{A}_{t,n}A_{t,n}^3}\left(Y_{t,n}\bs{W}_{t,n}^{H} + \overline{Y}_{t,n}\bs{W}_{t,n}^{\top}\right)\left(\overline{\hat{Y}}_{t,n}Y_{t,n} + \hat{Y}_{t,n}\overline{Y}_{t,n}\right) + \frac{1}{\hat{A}_{t,n}A_{t,n}}\left(\overline{\hat{Y}}_{t,n}\bs{W}_{t,n}^{\top} + \hat{Y}_{t,n}\bs{W}_{t,n}^{H}\right)  \right) \label{eq:cos_pd_2} \\
&= \frac{1}{2} \left(-\frac{1}{2\hat{A}_{t,n}A_{t,n}^3}\left(\overline{\hat{Y}}_{t,n}Y_{t,n}^2\bs{W}_{t,n}^{H} + \hat{Y}_{t,n}A_{t,n}^2\bs{W}_{t,n}^{H} + \overline{\hat{Y}}_{t,n}A_{t,n}^2\bs{W}_{t,n}^{\top} + \hat{Y}_{t,n}\overline{Y}_{t,n}^2\bs{W}_{t,n}^{\top} \right) + \frac{1}{\hat{A}_{t,n}A_{t,n}}\left(\overline{\hat{Y}}_{t,n}\bs{W}_{t,n}^{\top} + \hat{Y}_{t,n}\bs{W}_{t,n}^{H}\right) \right) \label{eq:cos_pd_3}\\
&= \frac{1}{2}\left(-\frac{1}{2\hat{A}_{t,n}A_{t,n}}\left(\overline{\hat{Y}}_{t,n}\frac{Y_{t,n}}{\overline{Y}_{t,n}}\bs{W}_{t,n}^{H} + \hat{Y}_{t,n}\bs{W}_{t,n}^{H} + \overline{\hat{Y}}_{t,n}\bs{W}_{t,n}^{\top} + \hat{Y}_{t,n}\frac{\overline{Y}_{t,n}}{Y_{t,n}}\bs{W}_{t,n}^{\top}\right) + \frac{1}{\hat{A}_{t,n}A_{t,n}}\left(\overline{\hat{Y}}_{t,n}\bs{W}_{t,n}^{\top} + \hat{Y}_{t,n}\bs{W}_{t,n}^{H}\right) \right) \label{eq:cos_pd_4}\\
&= \frac{1}{2} \left(\frac{1}{2\hat{A}_{t,n}A_{t,n}} \left(\hat{Y}_{t,n}\bs{W}_{t,n}^{H} - \hat{Y}_{t,n}\frac{\overline{Y}_{t,n}}{Y_{t,n}}\bs{W}_{t,n}^{\top} + \overline{\hat{Y}}_{t,n}\bs{W}_{t,n}^{\top} - \overline{\hat{Y}}_{t,n}\frac{Y_{t,n}}{\overline{Y}_{t,n}}\bs{W}_{t,n}^{H}\right) \right) \label{eq:cos_pd_5}\\
&= \frac{1}{4} \left(\frac{\hat{Y}_{t,n}}{\hat{A}_{t,n}}\frac{\overline{Y}_{t,n}}{A_{t,n}}\left(\frac{1}{\overline{Y}_{t,n}}\bs{W}_{t,n}^{H}-\frac{1}{Y_{t,n}}\bs{W}_{t,n}^{\top}\right) + \frac{\overline{\hat{Y}}_{t,n}}{\hat{A}_{t,n}}\frac{Y_{t,n}}{A_{t,n}}\left(\frac{1}{Y_{t,n}}\bs{W}_{t,n}^{\top} - \frac{1}{\overline{Y}_{t,n}}\bs{W}_{t,n}^{H} \right) \right) \label{eq:cos_pd_6}\\
&= \frac{1}{4} \left(\frac{\hat{Y}_{t,n}}{\hat{A}_{t,n}}\frac{A_{t,n}}{Y_{t,n}}\left(\frac{1}{\overline{Y}_{t,n}}\bs{W}_{t,n}^{H}-\frac{1}{Y_{t,n}}\bs{W}_{t,n}^{\top}\right) + \frac{\overline{\hat{Y}}_{t,n}}{\hat{A}_{t,n}}\frac{A_{t,n}}{\overline{Y}_{t,n}}\left(\frac{1}{Y_{t,n}}\bs{W}_{t,n}^{\top} - \frac{1}{\overline{Y}_{t,n}}\bs{W}_{t,n}^{H} \right) \right) \label{eq:cos_pd_7}\\
&= \frac{1}{2} \mathcal{R}\left(\frac{\hat{Y}_{t,n}}{\hat{A}_{t,n}}\frac{A_{t,n}}{Y_{t,n}}\left(\frac{1}{Y_{t,n}}\bs{W}_{t,n}^{\top}-\frac{1}{\overline{Y}_{t,n}}\bs{W}_{t,n}^{H}\right)\right) \\
&= \mathcal{I}\left(\exp(i(\angle \hat{Y}_{t,n} - \angle Y_{t,n}))\right)\mathcal{I}\left(\frac{1}{\overline{Y}_{t,n}}\bs{W}_{t,n}^{H}\right) \\
&= \sin(\angle \hat{Y}_{t,n} - \angle Y_{t,n}) \mathcal{I}\left(\frac{1}{\overline{Y}_{t,n}}\bs{W}_{t,n}^{H}\right).
\end{align}

\end{document}